\documentclass{IEEEcsmag}

\usepackage[colorlinks,urlcolor=blue,linkcolor=blue,citecolor=blue]{hyperref}

\usepackage{upmath}
\usepackage{graphicx}
\graphicspath{ {./images/} }

\jvol{XX}
\jnum{XX}
\paper{8}
\jmonth{May/June}
\jname{IT Professional}
\pubyear{2021}

\begin{document}

\sptitle{Department: Head}
\editor{Editor: Name, xxxx@email}

\title{DiVRsify: Break the Cycle and Develop VR for Everyone}

\author{Tabitha C. Peck}
\affil{Davidson College}

\author{Kyla McMullen}
\affil{University of Florida}

\author{John Quarles}
\affil{The University of Texas at San Antonio}

\markboth{Department Head}{Paper title}

\begin{abstract}
Virtual reality technology is biased. It excludes approximately 95\% the world's population by being primarily designed for male, western, educated, industrial, rich, and democratic populations. This bias may be due to the lack of diversity in virtual reality researchers, research participants, developers, and end users, fueling a noninclusive research, development, and usability cycle. The objective of this paper is to highlight the minimal virtual reality research involving understudied populations with respect to dimensions of diversity, such as gender, race, culture, ethnicity, age, disability, and neurodivergence. Specifically, we highlight numerous differences in virtual reality usability between underrepresented groups compared to commonly studied populations. These differences illustrate the lack of generalizability of prior virtual reality research.  Lastly, we present a call to action with the aim that, over time, will break the cycle and enable virtual reality for everyone.
\end{abstract}

\maketitle

\chapterinitial{Virtual Reality} (VR) researchers argue that the diverse array of VR applications has the potential to change the world. A subset of these applications includes driving and flight simulators, surgical training, exposure therapy, physical therapy, empathy exercises, and perspective taking. Although these applications are intended to be useful for everyone, regardless of their age, gender, race, culture, ethnicity, class, ability, neurodiversity, etc., the majority of VR development is focused on a minority of the population. This narrow focus limits what we are calling the VR research, development, and usability cycle. This cycle excludes the majority of the population from being involved in the use, research, and development of VR applications and hardware.  

\begin{figure}[t]
\includegraphics[width=7cm]{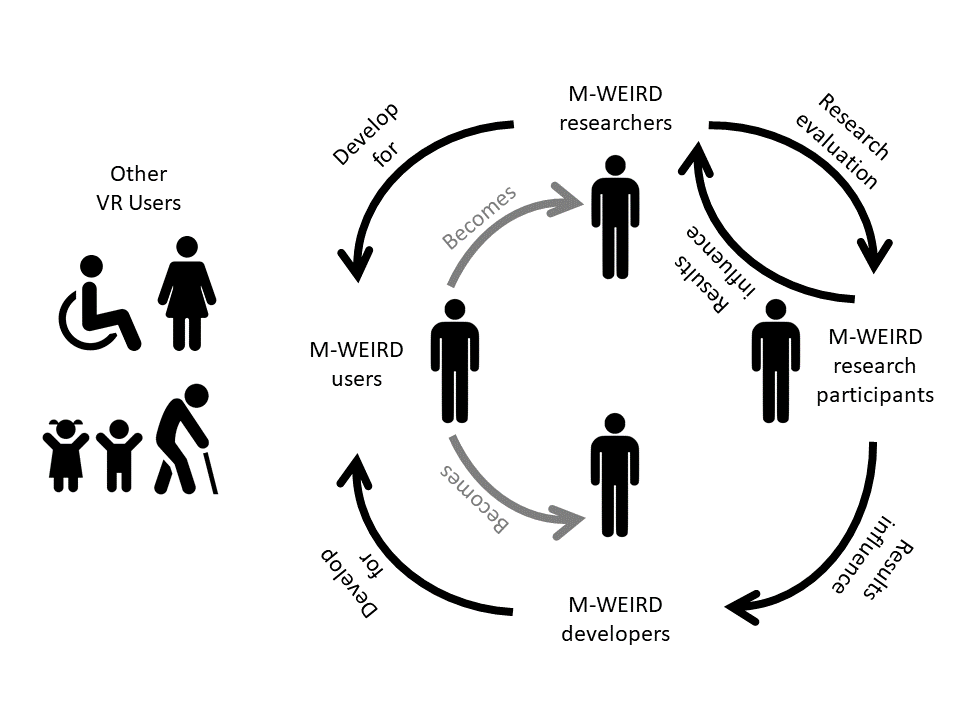}
\caption{The noninclusive VR research, development, and usability cycle.}
\label{fig:cycle}
\end{figure}

The VR research, development, and usability cycle (\textbf{Figure \ref{fig:cycle}}) begins with the majority of researchers being \textbf{M}ale,  \textbf{W}hite, \textbf{E}ducated, \textbf{I}ndustrialized, \textbf{R}ich, and \textbf{D}emocratic (M-WEIRD) \cite{henrich2010weirdest}. The WEIRD population is less than 6\% of the worlds population. Assuming that men are roughly half of this population the M-WEIRD population is representative of less than 3\% of the world's population. 
At their inception, VR research questions are primarily created by M-WEIRD researchers and evaluated on M-WEIRD participants \cite{peck2020mind}. Subsequently, the knowledge gained from this limited subsection of the population is broadly applied 1) in industry by  M-WEIRD developers and then 2) influences future research by M-WEIRD researchers. The hardware and software developed by industry professionals, based on knowledge gained and investigated by M-WEIRD researchers and participants is then used by M-WEIRD users. 
This cycle often excludes other VR users who do not identify as M-WEIRD and may limit the broad effectiveness of VR research and applications. 

 Research and design decisions may disproportionately dissuade non-M-WEIRD people from effectively using VR, due to VR research and development primarily focusing on the M-WEIRD population. For example, modern head-mounted displays (HMDs) used for perceiving VR do not accommodate women or children. The interpupillary distance (IPD) accommodated in modern HMDs supports 81\% of men's IPDs yet only 61\% of women's and even fewer children's \cite{dodgson2004variation}. 
 This mismatch in IPD can cause discomfort and eyestrain 
 as well as in depth perception errors. 
  This discrepancy may also explain why the Oculus Rift was found to be more nauseating for women than men \cite{munafo2017virtual}. In addition, the physical design of HMDs for virtual reality is often unaccommodating of the hair and headdresses of marginalized racial and ethnic groups. 

Because recreational VR has primarily been designed for M-WEIRD users,  M-WEIRD users are more likely to have positive experiences with VR. These experiences may influence some M-WEIRD users to become VR researchers and developers. The authors have witnessed this anecdotally in the students who elect to enter the field of computer science due to their interest in video games and/or VR. Further, having systems designed or advertised for M-WEIRD users may propagate the lack-of-fit model \cite{heilman1983sex} and build a barrier to entry for under-served populations using VR. Regardless of how or why people decide to become developers and researchers, it is clear that the people in this cycle are majority M-WEIRD.  This exclusionary practice continues the cycle where VR will be developed for and designed by a non-diverse group of people. This cycle may inadvertently exclude, limit usability, or dissuade potential VR users that do not identify as M-WEIRD from using VR or becoming researchers or developers. 

The VR research, development, and usability cycle is of course simplified for illustration purposes and does not take into account additional external factors.  These other factors may affect who uses and develops VR hardware and software, and include societal and family pressures, economic barriers, stereotype threats, exclusionary messaging, etc. 

Immediately creating a more diverse group of researchers and developers may not be possible, however building a diverse group of researchers and developers is an important future goal. Further, asking a more diverse group of users to quickly adopt VR may not initially result in wide acceptance, especially if applications and hardware are not designed to support these users. Nevertheless, the VR research community does have the ability to enforce participant population diversity to include a wider range of participants and to ask research questions that consider the needs of all potential users. Having researchers and reviewers commit to the simple act of diversifying participant populations may have profound impacts on research results and break the current, noninclusive VR research, development, and usability cycle so that future VR is designed to be more inclusive.

The goal of this article is to highlight the importance of inclusivity when creating VR hardware, software, and applications. There is a need to include more diverse populations in the VR research, development, and usability cycle. This work introduces the negative impact of design bias  and presents research identifying the differences between different demographic groups within VR research. Finally, a  call to action is presented for the VR research community to improve our current practices, with the goal of making VR more accessible for everyone. 

\section{DESIGN BIAS}
\textit{Bias} is defined as an unequal weighting in favor of one group compared to another.\textit{ Design bias} is ``the development and dissemination of hardware and software whose characteristics systematically do not meet the needs of a subset of target users'' \cite{peck2020mind}. Design bias can reduce the generalizability of research study results and can negatively impact the usability and accessibility of designs. 
 To mitigate bias, VR researchers and developers should include feedback from more diverse populations when designing VR hardware and software. One way to include diverse feedback is through \textit{user-centered design}, a design technique used to engage and involve critical stakeholders in the co-design of the technologies they use. Ensuring a diverse participatory design population can help to reduce design bias.
This involvement can occur at any stage of the design process and includes methods such as focus groups, prototype development, and developing storyboards. Historically, design bias has resulted in negative consequences, as seen in the following cautionary example. 

The automotive industry historically excludes female passengers when evaluating vehicle safety. As recently as 2010, automotive safety was primarily tested using crash-test dummies that represented an average-sized, adult male body. Because female passengers were not represented during testing, in the same car accident women had a 47\% higher chance of injury compared to men \cite{bose2011vulnerability}. Even as of today there are no crash-test dummies based on female bodies, rather they are scaled-down male dummies, which still increases risk of injury for female passengers.

Design bias has also affected many underrepresented groups, such as women, racial and ethnic minorities, children, the elderly, transgender people, people with physical disabilities, and people with cognitive disabilities.
These types of biases can appear in physical hardware or software contexts and have been observed in a variety of scenarios from facial recognition to video games.

It is generally agreed that bias can be divided into three categories: preexisting bias, technical bias, and emergent bias \cite{friedman-1996}. \textit{Preexisting bias} occurs when the designer (knowingly or unknowingly) infuses their own inherit biases into the product they are creating. For example, this can occur in VR, during visuomotor synchrony, when the virtual world is designed such that all users must embody a White virtual avatar. \textit{Technical bias} occurs when the system design limits the usability. This can occurs when designing an interface for only right-handed users, or designing a head-mounted display that does not support all user's IPDs. \textit{Emergent bias} occurs over time and as people use the system. It can occur during development or after product release. Consider the case of virtual characters learning verbal interactions driven by machine learning (ML). Using biased training data, virtual characters have learned sexist and racist behaviors \cite{neff2016talking}.

\section{PARTICIPANT UNDERREPRESENTATION}
 VR research often requires data collected from human participants, similar to many other research communities. Many of these research communities, including ML, human-computer interaction, and the medical industry have called for using more diverse and representative participant populations.
 Some of these calls to action have stemmed from research highlighting the problems of underrepresentation in participant samples in psychology research.

 Similarly to VR, most research in psychology is based on participants who are WEIRD \cite{henrich2010weirdest}.  Henrich et al. \cite{henrich2010weirdest} highlight several instances where non-WEIRD study participants yield very different results compared to WEIRD participants, even in studies on aspects of human perception that were once thought to be universal. For example, consider the Mueller‐Lyer Illusion, which is a perceptual illusion where two lines are judged to be different lengths based upon the orientation of the arrows at the ends. Results of prior studies have shown this to be a very strong illusion in WEIRD societies, but many non-WEIRD societies are not fooled by this illusion. This suggests that since human visual perception varies  greatly over such a simple visual illusion, then visual perception differences are  likely to be found in VR as well. 

 Similar concerns have been raised about using non-representative samples in ML data sets. 
 It is widely agreed that when training a machine learning algorithm, choosing appropriate and representative training data is critical. If the training data is biased, then then ML algorithm predictions will also be biased. The majority of the ML literature tends to focus on sample selection that only includes observable data, meaning that the data that is chosen for training only includes the data that could be directly observed. Oftentimes the predicted outcomes could be influenced by unobservable or uncollected data. For example, suppose someone wanted to create an algorithm to predict PhD student retention using data from past PhD graduates. We would only observe the outcome of a student finishing a PhD program if the student decided to stay in the PhD program. The observed outcomes are only a consequence of a human-decision maker. There are unobservable factors (such as family dynamics, advisor relationship, financial burden, stereotype threat, department climate, experienced microaggressions etc.) that influence the outcome. This scenario makes it challenging to create a predictive model since the outcomes observed do not represent a random sampling of all PhD students who entered the program (not just the ones who received their PhD). Oftentimes in ML, the mechanism that determines selection has no impact on the outcome being conditional on the observed attributes. 
 Standard ML classifiers assume that data is drawn independently and normally distributed, however the selection of examples is often biased, thus leading to biased inferences. For example, training sets consisting of a majority of light-skinned and male faces are speculated to be the reason why face-recognition systems are most accurate at identifying light-skinned male faces, and are less accurate at identifying female faces and dark-skinned faces \cite{buolamwini2018gender}. Using ML within VR is becoming more common, and this trend further highlights the importance of using diverse and representative data sets for training in all fields. 
 
 Finally, Peck et al. \cite{peck2020mind} demonstrated that female participants were significantly underrepresented in VR research and that this underrepresentation biased research findings. The change in simulator sickness after VR exposure was systematically proportional to the number of women who participated in experiments such that experiments with more women had a smaller increase in simulator sickness. If simulator sickness results were systematically related to gender it can be assumed that other unknown measures may also be affected. This further demonstrates the importance of using diverse participant populations when performing VR research.

\section{DIVERSITY DIMENSIONS}
Even though numerous research communities have called for including more diverse participant samples, it is unclear which diversity dimensions are important to consider. The WEIRD dimensions of race/ethnicity, education, industrialization level, socioeconomic status, and government type are only a small subset of the dimensions of human diversity.
Moreover, this subset of dimensions is not regularly reported in research papers, which often report little more than age and binary-gender. These additional diversity dimensions include, but are not limited to age, culture, gender, sex, mental abilities, physical abilities, sexual orientation, appearance and body, class, geographic location, language and accent, migration biographies, parental status, relationship status, and religion.


To evaluate the inclusion of diversity dimensions in human computer interaction research, Himmelsbach et al. \cite{himmelsbach2019we} investigated how many papers in the ACM Conference on Human Factors in Computing Systems (CHI) included diversity dimensions and counted how many dimensions were included in reports of participant demographics. They found that there was a significant increase in the number of diversity dimensions reported per paper between 2001 and 2011. However, there was no significant difference between 2011 and 2016. The average number of dimensions reported in 2016 was approximately 3 out of 14 considered dimensions. Thus, it seems that human diversity has been largely ignored in the CHI community, which overlaps significantly with VR research. 

Determining the diversity dimensions that most influence usability and design remains an open research question. Collecting and reporting diversity dimensions will provide readers a better understanding of the participant populations and may provide insight into demographic groups that should be evaluated in further detail. 

\section{PARTICIPANT POPULATION DIFFERENCES\label{sec:differences}}
In this section we present numerous studies that identified response differences based on diversity dimensions. These differences could be caused by numerous factors including biological, behavioral, situational, or cultural differences. 
Although we present experiments identifying differences based on diversity dimensions, we do not hypothesize on what caused these identified differences. 

The authors acknowledge that there are differences within and between the diversity dimension groups identified in this section. As such, we recognize that there are unique effects due to intersectionality that cannot be captured by studying one dimension in isolation. Our hope is that by bringing awareness to the necessity of research along various identity dimensions, future research can explore intersectional experiences in VR. 

\subsection{Gender}
Numerous gender differences have been reported in VR research, however it is likely that many more differences have never been explored.  For example, studies specifically looking for gender differences have identified them in perceptual threshold studies. 

Additional differences have been found between genders in response to simulator sickness. However, results are mixed and suggest that women are more likely to get sick compared to men, 
have comparable levels of simulator sickness to men, 
or based on a 5-year meta analysis may be less likely to get simulator sickness compared to men \cite{peck2020mind}. This discrepancy suggests that gender differences in simulator sickness is still an open question, where additional factors such as environment type, participant age, or participant VR experience may need to be considered to fully understand response differences within VR experiences.

Additional unexpected gender differences have been identified in VR scenarios. Women and men used hand-held trackers differently, 
experienced different levels of spatial immersion, 
and women outperformed men on spatial performance tasks. 
Women and men experienced embodiment in a self avatar differently such that women were less likely to accept avatar hands that were different from their own appearance. 
Demonstrating preexisting bias, this effect was further seen in a lowered embodiment score for women compared to men when an average male-sized hand was used in an experiment \cite{peck2021avatar}.

Even though differences have been detected, more often than not, gender differences are rarely tested. When differences are tested, studies may only be sufficiently powered to detect large effects while gender differences are likely to be small effects \cite{peck2020mind}. The sample size in many studies is small (n $\leq$ 40) and disproportionately male, further complicating the investigation of potential differences.

\subsection{Race, Culture, and Ethnicity}
Additional diversity dimensions that are seldom evaluated and often conflated include race, culture, and ethnicity. As noted by Henrich et al. \cite{henrich2010weirdest} people who did not identify as WEIRD perceived a perceptual illusion differently than WEIRD people. This work suggests that differences may exist due to race, culture, and ethnicity.

 For example in VR research, Almog et al.\cite{almog2009ethnicity} conducted a study including Israeli,  Arab and non-Arab men and women. Participants rode in a virtual airplane.  The plane took off, flew in nice weather, flew in stormy weather, and then landed. Results suggest that Arab women were significantly less likely than non-Arab women to look out the window of the airplane. This behavior was significantly correlated with their self-reported sense of presence.

Additionally, Olson \cite{olson2019exploring} performed a qualitative analysis of how participants' experience growing up with racial and ethnic socialization (RES) affected decisions in a VR game including racial discrimination themes. The game, Passage Home VR, was an interactive narrative where the player is accused of plagiarism in an educational setting. The body language of the avatar determined the events in the narrative. The author found that participants' previous experiences with RES informed their decisions in VR. 

The limited work in this area may be due to different interpretations of race, culture, and ethnicity worldwide as well as the challenge of creating a universal questionnaire  for collecting this information. Peck et al. \cite{peck2020mind} proposed a questionnaire and recommended allowing participants to self-identify. Additionally, many research locations may not be racially, ethnically, or culturally diverse, thus limiting the option of investigating differences along these dimensions.  Even though there are obvious challenges for investigating race, culture, and ethnicity it is still possible to collect this participant data and to report it in papers. 
 
\subsection{Age}

Though VR may be created with an adult end-user in mind, attention should be paid to VR applications for populations of any age. For example, virtual reality has positively impacted children and older adults by improving physical activity.  VR games increase physical activity in children 
and research has demonstrated that older adults who use VR for exercise increase their mobility and decrease their likelihood of falling. 
Conversely, older populations may also be more reluctant to use VR for a myriad of reasons such as user expectations, demographic segmentation in marketing, and lack of familiarity.Future efforts should aim to mitigate these issues to promote the inclusion of older adults.

Participant age can also affect presence and embodiment in VR. For example, Mcglynn et al. \cite{mcglynn2019investigating} investigated the influence of age on participants' sense of presence. Participants played a VR game called 'Diner Duo' in which they had to make and serve virtual hamburgers to virtual customers in a virtual diner. Results suggest that spatial presence was not significantly different across ages but older participants were less likely to notice breaks in presence.

When considering embodiment, Serino et al. \cite{serino2018role} investigated the influence of age in body size perception in VR. Participants experienced an embodiment illusion with an avatar that had smaller proportions than the participants. One task for the participants was to estimate the width of their hips both before and after VR exposure. The 19-to-25-year-old participants increasingly underestimated hip width post VR exposure, whereas the 26-to-55-year-old participants width estimations were not significantly affected by VR. Further, Peck and Gonzalez-Franco \cite{peck2021avatar} identified that participants over 30 had a lower sense of embodiment in a self-avatar compared to participants under 30.

Children are a protected population and although VR may be beneficial in numerous applications including increasing physical activity, 
pain distraction, 
or education 
care must be taken when developing and designing for this group. Children are physically smaller than adults and current HMDs are not designed for children and do not support their smaller IPDs.
Further, children are actively in the process of cognitive development and may respond in unanticipated ways when put into virtual environments. 

For example, Segovia et al.  \cite{segovia2009virtually} investigated children's acquisition of false memories as an effect of VR exposure. A false memory in VR implies believing that what happened in the VE happened instead in the real world. They compared several different conditions that could impact memory - idle, mental imagery, VR with another child's avatar, and VR with a self avatar. Results suggested that preschool children were equally likely to acquire false memories in any condition. However, elementary school children acquired more false memories in the mental imagery and VR self-avatar conditions than the idle condition. This finding suggests that VR may have different effects on users' memory depending on age.

\subsection{Disability}
VR is not accessible to many people with disabilities. Unfortunately, these populations are rarely consulted during the non-inclusive VR research, development and usability cycle, which creates a major barrier to accessibility. For example, some people with balance impairments may not be able to safely stand up in many standing-based VR experiences. This limitation may prevent users with balance impairments from engaging in all parts of the experience. 
Oftentimes, disabled people are only considered in VR research for the sole purpose of rehabilitation or correction applications. VR designers never initiate research to study recreational uses of VR for disabled people and instead treat them as a deviation from "normal" which needs correction \cite{spiel-2021}. 

There have been various studies that have investigated the unique experiences and responses that persons with disabilities have in VR. For example, numerous studies support the claim that people with Multiple Sclerosis (MS) have a different experience of presence than people without MS. 
\cite{guo2014effects}.
Moreover, persons with MS may respond to latency differently than persons without disabilities\cite{RN248}. 

Additional studies have focused on the experience of cybersickness on people with MS as compared to people without MS, specifically looking at physiological and brain wave responses.Many of the baseline differences persisted in VR. However, in some cases
VR induced completely opposite changes between the participants with MS and participants without MS. Differences were found near the parietal lobe - sensation, perception, integration of sensory input and decision making, emotional behavior, and visual reception - and the frontal lobe, which mainly deals with motor functions\cite{RN347}. These results suggest VR may need to be made more adaptive and accessible to the needs of persons with disabilities.

\subsection{Neurodiversity}
VR is widely accepted as a training tool for learning various skills, however, this may not be an acceptable intervention for neurodiverse people. In fact, in some instances VR has been harmful in practice when applied to autism intervention. Williams and Gilbert \cite{williams-2020} conducted a survey of wearable technologies applied to autism intervention. Their work found that 90\% of interventions focused on "normalizing" autistic people, viewing their traits as deficits. Only 10 \% of the technologies surveyed addressed user needs for sensory regulation, emotional regulation, communication, or executive function. It is critical to ensure that VR applications for neurodiverse people do not follow this pattern. 

In the VR domain, Self et al. \cite{self-2007} created a VR learning environment for children with autism spectrum disorder to learn fire safety skills. All of the children in the study learned and demonstrated an understanding of fire safety skills in the virtual environment. However, these results did not always generalize to the real world. For example, when a fire alarm was triggered in the real world, several students with ASD still needed verbal prompting to exit. Moreover aspects that can affect learning in VR such as interfaces, may not be inclusive to neurodivergent people. Mei et al. \cite{RN137} investigated   how   children   with   autism spectrum disorder performed basic 3D interaction  tasks,  such  as  rotation  and  translation, compared to typically developed  children. Participants had two  tasks:  1)  rotating  a  virtual  object  to match the pose of a target object, and 2) translating  a  virtual  object  to  the same relative position as a target object. Results suggested that on tasks requiring attention to  precision  (e.g.,  translation  along  the  z  axis)  and  tasks  in  which  the  interface  was  more  ergonomically  limiting  (e.g.,  rotation  tasks  when  trying  to  align  the  controller  with the virtual object), the autism spectrum disorder group demonstrated significantly longer completion time and error prone performance than  the  typically developed  group. These findings suggest that additional research should seek to understand how neurodiverse populations learn and apply skills, such that VR can adapt to these specific needs.  

\section{A CALL TO ACTION}
The previous sections presented examples of significant differences in how participants respond to VR based on different diversity dimensions. In some of the above examples the initial research goal was not to investigate differences in diversity dimensions. For example, the difference in the subjective sense of embodiment by age was identified because of a diverse participant sample even though this was not the intended goal of the study \cite{peck2021avatar, peck2021evidence}. This example highlights the importance of diversifying sample populations when designing for general populations instead of relying primarily on convenience sampling of M-WEIRD college-aged participants. When working with untested populations, researchers may find unanticipated results that lead to new lines of research and generate more generalizable results across diverse populations. 

We argue the the VR research community is ethically obligated to develop inclusive VR hardware and applications. To support this, we propose several actions that every VR researcher can take to improve the generalizability of their research through engaging with more diverse participant populations and researchers:

\textbf{Place greater emphasis on population diversity.} When reviewing papers, consider if the participant population is representative of the intended population for the proposed research. If the studied population lacks appropriate diversity (gender, age, race, etc.) provide constructive criticism and require the paper to highlight this limitation and accurately quantify the results according to the lack of participant diversity. Conference and journal review criteria should include evaluation of participant diversity and place more weight on generalizability of results such that papers with greater participant diversity, or research including understudied populations are more likely to be published.

\textbf{Actively recruit participants from outside your university.} Recruiting primarily from university students is a common method of conducting studies in psychology, HCI, and VR, because it is convenient. However, the diversity of university students is limited to a narrow range of ages and educational backgrounds. Thus, sampling only university students may not be sufficient for generalizability. Therefore,  most other populations can be considered underrepresented in the scope of VR research. 

A first step in the right direction would be to recruit from the local  population. For example, recruiting through online social media platforms, like Reddit, working with local businesses to hang flyers in store windows, or collaborating with local affinity groups. The people recruited through these means should be paid for their contribution, including any financial costs incurred from traveling to the laboratory to conduct the experiment (i.e.- gas, parking, compensation for long travel time). If possible, you should try to bring the experiment to these outside people, rather than requiring them to travel to you.

\textbf{Re-evaluate your evaluation methods.}
Differences may be observed between various populations in VR research, however it is critical to understand if these differences are characteristic of the population, or if some aspect of the methodology is affecting performance. In addition to collecting quantitative experimental data, qualitative data analyzing the participants' experiences should be taken. This information will be critical in making the experiment design more inclusive.
A first step in this direction would be to include an exit survey at the conclusion of each experiment, asking open ended questions about the user's experience, how they perceived the virtual world, and any factors that may have helped or hindered their performance.

\textbf{Use inclusive imagery in recruitment materials, environmental scenes, and publication images.} Imagery provides a subtle cue as to who is ``welcome" in a space. For example, a recruitment flyer with an image of a White man may subtly indicate that women and non-White people do not belong. Additionally, hyper-sexualized images of women are non-professional and propagate hostile unwelcoming environments. Make sure the imagery is inclusive during recruitment and publication and use it to encourage the widest variety of people to participate. Show recruitment flyers to a diverse group during the design process; listen and act if someone finds the materials to be offensive or non-inclusive. When creating materials, be aware of and avoid stereotypes.

\textbf{Develop a relationship with affinity groups.} Affinity groups are groups of people that meet and have interests. Examples include include women in the workplace, working parents, lesbian, gay, bisexual, transgender, queer or questioning etc. (LGBTQ+) affinities. A positive and collaborative relationship can increase the diversity of the project's research participants or provide diverse insights and perspective on research projects. When working with affinity groups it is critical that you inform the group of your intentions, for example to increase recruitment diversity or in an advisory role. Be mindful that the relationship should be mutually beneficial. You could commit to recruiting research students within this group, giving research presentations, or applying for funding to support their advisory role.

\textbf{Collaborate with researchers from diverse geographic populations.} While recruiting from the local general population is a step in the right direction towards increasing participant diversity, it is not sufficient. The better method of increasing participant diversity is to also recruit from beyond the local population. The easiest way to do this is to collaborate with researchers outside your institution or outside your country. Compatibility of hardware and software makes this research more challenging, however the development of commercialized VR hardware over the past decade makes collaboration between labs more feasible. Moreover, commercial VR systems are possible to ship for sharing with distant collaborators and participants.

\textbf{Replicate experiments and include underrepresented populations.} The replication crisis \cite{pashler2012editors} highlights the importance of replicating previous studies and valuing replication work. Replicating previous experiments and including diverse participant populations will determine if the experiment is replaceable, supports generalizability of results, or highlights usability differences between groups. Further, when evaluating papers, place a higher value on replication work that includes or investigates previously under-served populations.

\textbf{Collect and report participant diversity data.} When creating your demographic survey collect additional diversity dimension data and report these data in the participant section of your research papers. This should also be reflected in the discussion and limitations sections, either as a strength or weakness of the work. Adding information about the participant population informs other researchers about the generalizability of the data. It may also highlight interesting and unexpected differences between groups and guide further research. Administer the demographic survey after the experiment so as not to induce unexpected stereotype threats and always present inclusive choices in questions. 
For example, do not restrict gender to a binary response.

\textbf{Actively and consistently consider the perspective of someone much different than yourself.} In the design process, encourage the research team to ponder, how would someone of a different \textit{race}, \textit{educational background}, \textit{gender}, \textit{age}, \textit{ability}, \textit{ethnicity}, \textit{literacy level}, \textit{culture}, \textit{class}, \textit{language} experience this system? Would there be any barriers to their engagement? How can we mitigate those challenges? When in doubt, consult with someone from that demographic. Additionally, include people from a wide variety of demographics in an advisory capacity for VR projects to help ponder and answer these questions. If your research and development team does not have the necessary diversity, hire experts from those populations to help inform the trajectory of the research. 

\textbf{Diversify the people in charge.} Having diverse reviewers, program committees, keynote speakers, and awardees signifies that diversity is an important contribution worth acknowledging and rewarding since different perspectives enhance research quality. When diversifying committees, be aware that underrepresented populations are often asked to complete higher shares of advisory work. Respect and acknowledge their time and contributions. Do not recruit non M-WEIRD personnel solely to increase a diversity quota, but to instead listen, learn, and respect their differing opinions. Further, call people out if someone questions the accomplishments of someone based on their identity, and instead highlight their well-deserved accomplishment and the barriers they may have overcome along the way.  

\textbf{Continually Engage in Professional Development Focused on Diversity.} Caring about diversity is the first step. To better understand, recognize, and respect the importance of diversity requires professional development. Commit to educate yourself in this nuanced and intricate science and treat it with the same respect you would a new academic area. This professional development is an important part of developing as researchers to improve and develop VR for a diverse and inclusive society. Pick up a book, join a reading group, listen to a podcast, or attend a lecture. Search for available resources in your area and utilize them to make yourself a better researcher.  


Regardless of your experience level or background,  take the first step and commit to any one of the above actions to help break the noninclusive VR research, development, and usability cycle. If everyone takes one small step, we can start to dismantle the system and create VR experiences that are useful for and inclusive of everyone.

\section{ACKNOWLEDGMENT}

The authors would like to thank the following people for graciously providing their thoughts and insight on this manuscript: Dr. Rua Williams and Dr. Jessica J. Good.

\bibliographystyle{abbrv-doi}

\bibliography{refs}

\begin{thebibliography}{10}

\bibitem{almog2009ethnicity}
I.~Almog, H.~S. Wallach, and M.~P. Safir.
\newblock Ethnicity and sense of presence in a virtual environment: Arab
  women-a case in point.
\newblock In {\em 2009 Virtual Rehabilitation International Conference}, pp.
  78--82. IEEE, 2009.

\bibitem{RN347}
I.~M. Arafat, S.~M.~S. Ferdous, and J.~Quarles.
\newblock Cybersickness-provoking virtual reality alters brain signals of
  persons with multiple sclerosis.
\newblock In {\em 2018 IEEE Conference on Virtual Reality and 3D User
  Interfaces (VR)}, pp. 1--120. IEEE.

\bibitem{bose2011vulnerability}
D.~Bose, M.~Segui-Gomez, ScD, and J.~R. Crandall.
\newblock Vulnerability of female drivers involved in motor vehicle crashes: an
  analysis of us population at risk.
\newblock {\em American journal of public health}, 101(12):2368--2373, 2011.

\bibitem{buolamwini2018gender}
J.~Buolamwini and T.~Gebru.
\newblock Gender shades: Intersectional accuracy disparities in commercial
  gender classification.
\newblock In {\em Conference on fairness, accountability and transparency}, pp.
  77--91, 2018.

\bibitem{dodgson2004variation}
N.~A. Dodgson.
\newblock Variation and extrema of human interpupillary distance.
\newblock In {\em Stereoscopic Displays and Virtual Reality Systems XI}, vol.
  5291, pp. 36--46. International Society for Optics and Photonics, 2004.

\bibitem{friedman-1996}
B.~Friedman and H.~Nissenbaum.
\newblock Bias in computer systems.
\newblock {\em ACM Transactions on Information Systems (TOIS)}, 14(3):330--347,
  1996.

\bibitem{guo2014effects}
R.~Guo, G.~Samaraweera, and J.~Quarles.
\newblock The effects of avatars on presence in virtual environments for
  persons with mobility impairments.
\newblock In {\em Proceedings of the 24th International Conference on
  Artificial Reality and Telexistence and the 19th Eurographics Symposium on
  Virtual Environments}, pp. 1--8, 2014.

\bibitem{heilman1983sex}
M.~E. Heilman.
\newblock Sex bias in work settings: The lack of fit model.
\newblock {\em Research in organizational behavior}, 1983.

\bibitem{henrich2010weirdest}
J.~Henrich, S.~J. Heine, and A.~Norenzayan.
\newblock The weirdest people in the world?
\newblock {\em Behavioral and brain sciences}, 33(2-3):61--83, 2010.

\bibitem{himmelsbach2019we}
J.~Himmelsbach, S.~Schwarz, C.~Gerdenitsch, B.~Wais-Zechmann, J.~Bobeth, and
  M.~Tscheligi.
\newblock Do we care about diversity in human computer interaction: A
  comprehensive content analysis on diversity dimensions in research.
\newblock In {\em Proceedings of the 2019 CHI Conference on Human Factors in
  Computing Systems}, pp. 1--16, 2019.

\bibitem{mcglynn2019investigating}
S.~A. McGlynn.
\newblock {\em Investigating age-related differences in spatial presence
  formation and maintenance in virtual reality}.
\newblock PhD thesis, Georgia Institute of Technology, 2019.

\bibitem{RN137}
C.~Mei, L.~Mason, and J.~Quarles.
\newblock Usability issues with 3d user interfaces for adolescents with high
  functioning autism.
\newblock In {\em Proceedings of the 16th international ACM SIGACCESS
  conference on Computers \& accessibility}, pp. 99--106. ACM.

\bibitem{munafo2017virtual}
J.~Munafo, M.~Diedrick, and T.~A. Stoffregen.
\newblock The virtual reality head-mounted display oculus rift induces motion
  sickness and is sexist in its effects.
\newblock {\em Experimental brain research}, 235(3):889--901, 2017.

\bibitem{neff2016talking}
G.~Neff.
\newblock Talking to bots: Symbiotic agency and the case of tay.
\newblock {\em International Journal of Communication}, 2016.

\bibitem{olson2019exploring}
D.~M. Olson.
\newblock {\em Exploring the role of racial and ethnic socialization in virtual
  reality (VR) narratives}.
\newblock PhD thesis, Massachusetts Institute of Technology, 2019.

\bibitem{pashler2012editors}
H.~Pashler and E.-J. Wagenmakers.
\newblock Editors’ introduction to the special section on replicability in
  psychological science: A crisis of confidence?
\newblock {\em Perspectives on psychological science}, 7(6):528--530, 2012.

\bibitem{peck2021avatar}
T.~C. Peck and M.~Gonzalez-Franco.
\newblock Avatar embodiment. a standardized questionnaire.
\newblock {\em Frontiers in Virtual Reality}, 1:44, 2021. doi: {{%
10\hspace{.1pt}\discretionary{.}{%
}{.}\hspace{.4pt}3389\discretionary{/}{%
}{/}frvir\hspace{.1pt}\discretionary{.}{%
}{.}\hspace{.4pt}2020\hspace{.1pt}\discretionary{.}{%
}{.}\hspace{.4pt}575943}}


\bibitem{peck2021evidence}
T.~C. Peck, J.~J. Good, and K.~Seitz.
\newblock Evidence of racial bias using immersive virtual reality: Analysis of
  head and hand motions during shooting decisions.
\newblock {\em IEEE Transactions on Visualization and Computer Graphics},
  27(5):2502--2512, 2021.

\bibitem{peck2020mind}
T.~C. Peck, L.~E. Sockol, and S.~M. Hancock.
\newblock Mind the gap: The underrepresentation of female participants and
  authors in virtual reality research.
\newblock {\em IEEE Transactions on Visualization and Computer Graphics},
  26(5):1945--1954, 2020.

\bibitem{RN248}
G.~Samaraweera, R.~Guo, and J.~Quarles.
\newblock Head tracking latency in virtual environments revisited: Do users
  with multiple sclerosis notice latency less?
\newblock {\em Visualization and Computer Graphics, IEEE Transactions on},
  PP(22.5):1630--1636, 2016. doi: {{%
10\hspace{.1pt}\discretionary{.}{%
}{.}\hspace{.4pt}1109\discretionary{/}{%
}{/}TVCG\hspace{.1pt}\discretionary{.}{%
}{.}\hspace{.4pt}2015\hspace{.1pt}\discretionary{.}{%
}{.}\hspace{.4pt}2443783}}


\bibitem{segovia2009virtually}
K.~Y. Segovia and J.~N. Bailenson.
\newblock Virtually true: Children's acquisition of false memories in virtual
  reality.
\newblock {\em Media Psychology}, 12(4):371--393, 2009.

\bibitem{self-2007}
T.~Self, R.~R. Scudder, G.~Weheba, and D.~Crumrine.
\newblock A virtual approach to teaching safety skills to children with autism
  spectrum disorder.
\newblock {\em Topics in Language disorders}, 27(3):242--253, 2007.

\bibitem{serino2018role}
S.~Serino, F.~Scarpina, A.~Dakanalis, A.~Keizer, E.~Pedroli, G.~Castelnuovo,
  A.~Chirico, V.~Catallo, D.~Di~Lernia, and G.~Riva.
\newblock The role of age on multisensory bodily experience: an experimental
  study with a virtual reality full-body illusion.
\newblock {\em Cyberpsychology, Behavior, and Social Networking},
  21(5):304--310, 2018.

\bibitem{spiel-2021}
K.~Spiel.
\newblock The bodies of tei--investigating norms and assumptions in the design
  of embodied interaction.
\newblock In {\em Proceedings of the Fifteenth International Conference on
  Tangible, Embedded, and Embodied Interaction}, pp. 1--19, 2021.

\bibitem{williams-2020}
R.~M. Williams and J.~E. Gilbert.
\newblock Perseverations of the academy: A survey of wearable technologies
  applied to autism intervention.
\newblock {\em International Journal of Human-Computer Studies}, 143:102485,
  2020. doi: {{%
10\hspace{.1pt}\discretionary{.}{%
}{.}\hspace{.4pt}1016\discretionary{/}{%
}{/}j\hspace{.1pt}\discretionary{.}{%
}{.}\hspace{.4pt}ijhcs\hspace{.1pt}\discretionary{.}{%
}{.}\hspace{.4pt}2020\hspace{.1pt}\discretionary{.}{%
}{.}\hspace{.4pt}102485}}


\end{thebibliography}

\begin{IEEEbiography}{Tabitha C. Peck}{\,} is an Associate Professor of Mathematics and Computer Science at Davidson College. Dr. Peck received her PhD in Computer Science from The University of North Carolina at Chapel Hill in 2010. Dr. Peck's research involves developing and testing usable virtual reality systems. Her current research includes investigation of the psychological implications of embodiment in self-avatars with the goal of using avatars to reduce and mitigate bias. Dr. Peck has been both a Journal and Conference Paper Program Chair for IEEE VR and a Journal Paper Science and Technology Chair for IEEE ISMAR. She is a Review Editor for Frontiers in Virtual Reality and an Associate Editor for Presence. She has received numerous honorable mentions and nominations for best paper awards at IEEE VR, and received a NSF CAREER award in 2020. Dr. Peck is an IEEE Senior Member. Contact her at tapeck@davidson.edu.
\end{IEEEbiography}

\begin{IEEEbiography}{Kyla A. McMullen}{\,} is an Assistant Professor of Computer and Information Science and Engineering at the University of Florida. Dr. McMullen received her PhD in Computer Science from the University of Maryland, Baltimore County. Dr. McMullen directs the Sound\textbf{PAD} Lab, which focuses on the \textbf{P}erception, \textbf{A}pplication, and \textbf{D}evelopment of 3D audio technologies for virtual and augmented reality. Broadly, her research aims to elucidate the human and computational factors that researchers must consider in the design and use of 3D audio systems. Dr. McMullen has served as the General Chair for the International Community on Auditory Display (ICAD) Conference and the National Society of Blacks in Computing (NSBC). She won the NSF CAREER Award in 2019. Contact her at drkyla@ufl.edu.
\end{IEEEbiography}

\begin{IEEEbiography}{John Quarles,}{\,}is an Associate Professor of Computer Science at The University of Texas at San Antonio and the Chief Technology Officer of MedCognition, Inc. Dr. Quarles received his bachelor’s degree in Computer Science from The University of Texas at Austin in 2004 and his PhD in Computer Engineering from the University of Florida in 2009. He has conducted research in virtual, mixed, and augmented reality, serious games, and 3D User interfaces. Much of his work has focused on the accessibility of these technologies for persons with disabilities. He has published numerous works in top conferences such as IEEE VR and top journals, such as IEEE TVCG. He has been awarded significant funding from the National Institutes of Health, the Department of Defense, and the National Science Foundation, including the prestigious NSF CAREER award in 2014. He was diagnosed with Multiple Sclerosis in 2004, resulting a variety of disabilities that have inhibited his use of virtual reality. He has the unique experience of being both a VR researcher and an end user with disabilities, serving to inform his chosen research focus.  Contact him at John.Quarles@utsa.edu.
\end{IEEEbiography}

\end{document}